\documentclass[10pt,twocolumn]{article}
\usepackage[utf8]{inputenc}
\usepackage{graphicx}
\usepackage{lipsum}
\usepackage{url}
\usepackage{caption}
\usepackage{titling}
\usepackage{csvsimple}

\newcommand \myabstract[2][.8]{
  \renewcommand\maketitlehookd{
    \mbox{}\medskip\par
    \centering
    \begin{minipage}{.5\textwidth}
      #2
    \end{minipage}}}

\title{Rendering Portals in Virtual Reality}
\author{Milan van Zanten\\\normalsize{University of Basel}\\\normalsize{milan.vanzanten@unibas.ch}}
\date{December 21, 2022}
\myabstract{
	\begin{center}
		\textbf{Abstract}
	\end{center}

	Portals have many applications in the field of computer graphics. Recently, they have found use as a way of artificially increasing the available space in a virtual reality (VR) environment. In this paper, we will cover a technique for making the transition through a portal unnoticeable to the user. Additionally, we will measure the performance impact of rendering portals in a test scene and provide some insight into possible optimisations.}

\begin{document}

\maketitle

\section{Introduction}
There are several uses for portals in computer graphics including determining the visibility of parts of a 3D scene\cite{lowe:2005}, dividing a scene up into separate areas, rendering a mirror surface, or as traversable portals that can be seen and moved through. These uses can be broadly categorised as either an optimisation technique or a rendering trick. The former two of the mentioned applications are used to determine geometry that can be ignored when rendering a scene and speed up the process. Mirror surfaces and traversable portals though are effects purposefully implemented in an application to benefit the experience.

There are already many implementations of traversable portals in media like video games or architectural visualisations\cite{aliaga:1997}. In this paper, we will focus on an application of traversable portals concerning the space limitations in a VR experience.

One of the main challenges when implementing VR applications is immersion, because errors in tracking and latency are noticed particularly strong\cite{abrash:2013}. In an effort to maximise immersion, most of VR has moved from seated experiences with movement limited to three degrees of freedom (just rotation) to ``room-scale'' tracking. In addition to the rotation of the VR headset, the user's position is tracked either via external devices with fixed positions or by cameras that analyse the surroundings and use computer vision algorithms to determine the position. With the added positional tracking, the six degrees of freedom allow the user to move around the room freely. The only limitation now is the available space.

To move around virtual worlds larger than the available space, several different methods have been developed\cite{boletsis:2017}. Examples include head-directed locomotion and point \& teleport. Indisputably though, the technique that preserves immersion the most is actual walking inside the real space.

\subsection{Impossible Spaces}
\begin{figure}[ht]
	\includegraphics[width=\linewidth]{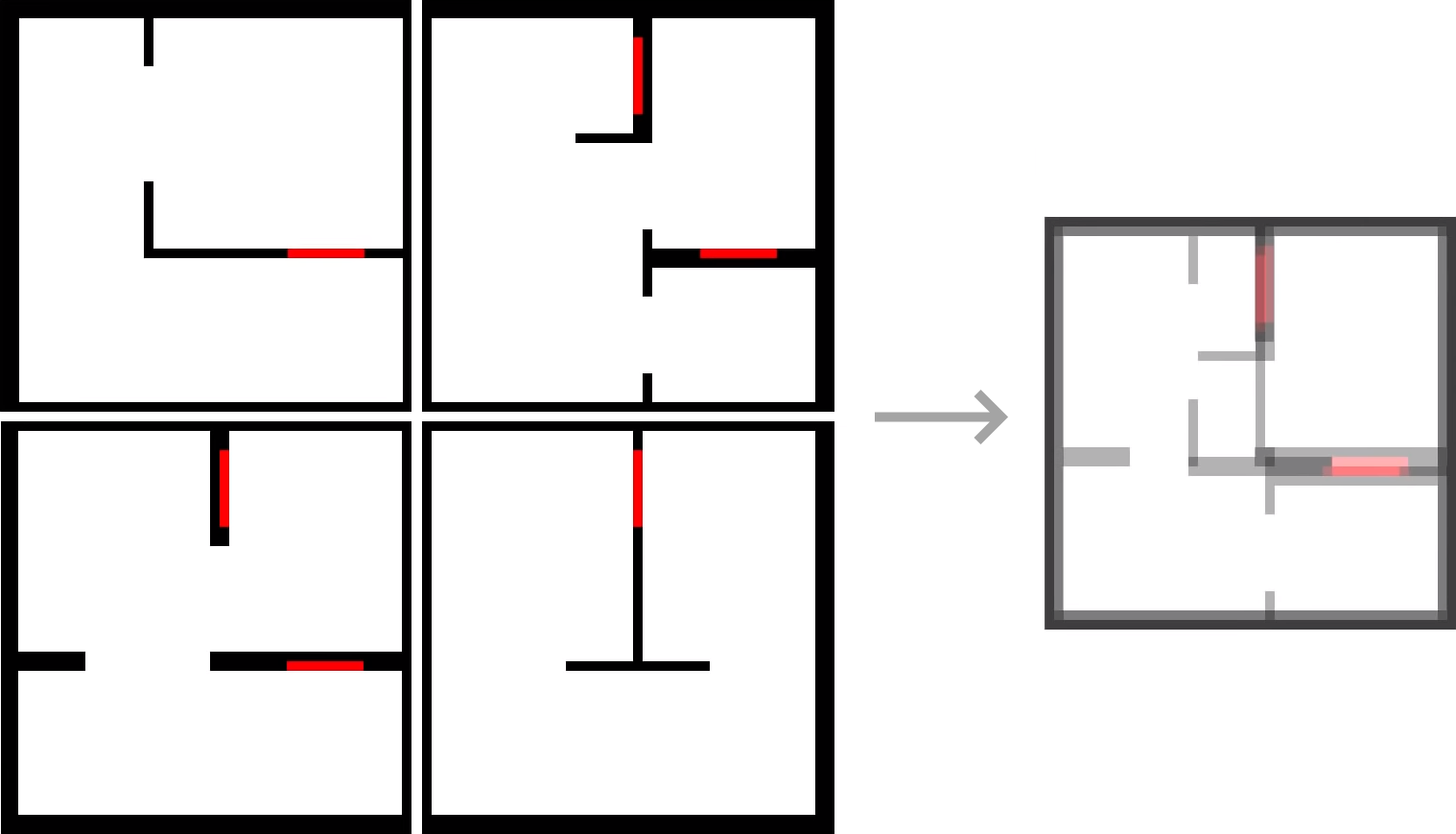}
	\caption{Four rooms with portals (in red) can all be accessed without leaving the smaller real space\cite{lochner:2021}.}
	\label{fig:impossible-spaces}
\end{figure}

A recent method to circumvent the space limitations of walking in a real space is the concept of impossible spaces. Overlapping rooms are connected through portals into a single space many times larger than the initial rooms themselves. If such an arrangement is made while factoring in the real available space, the whole virtual space can be accessed by passing through the portals. An example layout can be seen in Figure~\ref{fig:impossible-spaces}.

To allow for such impossible spaces to exist, the aforementioned portals are necessary. When viewed, they show what the user would be seeing through the portal in the other room, and if a user crosses the plane of a portal, they are transported to the connecting one.

In virtual reality, implementing such a portal system poses some extra challenges.

The goal of this paper is to explore how classical portals from monoscopic\footnote{An application that only shows one viewpoint is called monoscopic, whereas a VR application that renders one viewpoint per eye is called stereoscopic. A detailed explanation of the terminology can be found in Taştı and Avcı\cite{monoscopic-stereoscopic:2020}.} applications can be implemented in stereoscopic VR. We will focus on two main objectives:

\begin{enumerate}
	\item How can portals be implemented in a way that the user can look and pass through them without noticing?
	\item What performance pitfalls should be considered when rendering portals in VR and what optimisations can be applied to alleviate them?
\end{enumerate}

\section{Unobtrusive Transitions}
\label{unobtrusive-transitions}
Portals in computer graphics are generally just flat quads\footnote{A flat, rectangular object that can be used to show an image in a 3D environment.} onto which the view from within another portal is drawn. In monoscopic applications, a single camera moving from space $A$ through a portal into space $B$ will see a flat image of what is on the other side of it as long as it is still in space $A$. As soon as it crosses the portal plane, it is being transported to space $B$ and will render that space directly.

\begin{figure}[ht]
	\includegraphics[width=\linewidth]{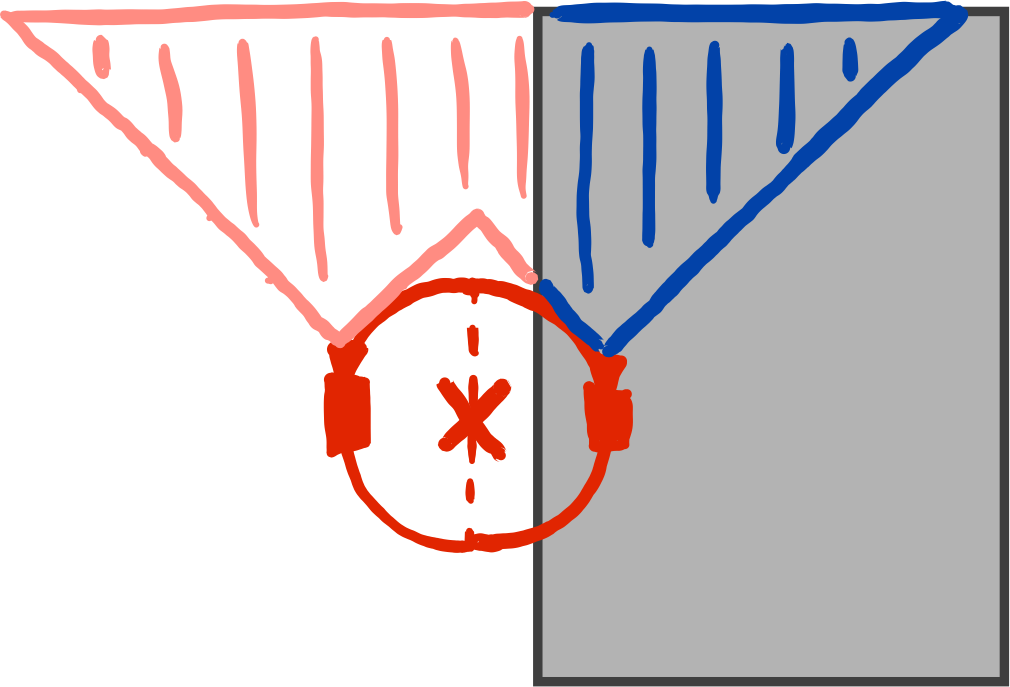}
	\caption{What should the right-eye camera render if it is inside the portal wall (in grey), but the centre of the head (in red) has not crossed the portal plane? If nothing is done, the blue part of the user's field of view would not render the next room, but whatever is inside or behind the portal wall.}
	\label{fig:obtrusive}
\end{figure}

The way a VR scene is rendered, though, raises the question of how to handle the teleportation of the two separate cameras rendering each eye without the user noticing. For example, consider the centre point between the eyes that could be used to decide when the portal plane has been crossed. If the user views the portal from an angle they could clip through the portal with one eye when they enter it before being transported, as shown in Figure~\ref{fig:obtrusive}.

At best, this results in a short flicker when the user passes through a portal. If the user stops inside the portal, however, it results in a completely wrong view on one eye.

One way to solve this problem could be to transport each eye separately whenever it passes through the portal, but that idea conflicts with one of the performance optimisations we will discuss in the next section.

If we want to keep the optimisations we introduce, a different method of making sure the transition through a portal is unnoticeable by the user, no matter how slowly they move or where they look. This method takes advantage of back-face culling\footnote{A triangle in computer graphics is generally considered to have a front side and a backside, determined by whether the points of the triangle appear in clockwise or counter-clockwise order from where they are viewed. Jeon et al.\cite{back-face-culling:2007} show how back-face culling reduces the amount of drawn triangles by only drawing the triangles that are facing the camera.}.

Instead of the portal being a plane, we will have it be a box. Every surface of the box is shaded with the view from the other room. If one eye clips inside the portal box, it will still see the other sides of the box that show the next room, instead of what was behind the portal originally.

\begin{figure}[ht]
	\includegraphics[width=\linewidth]{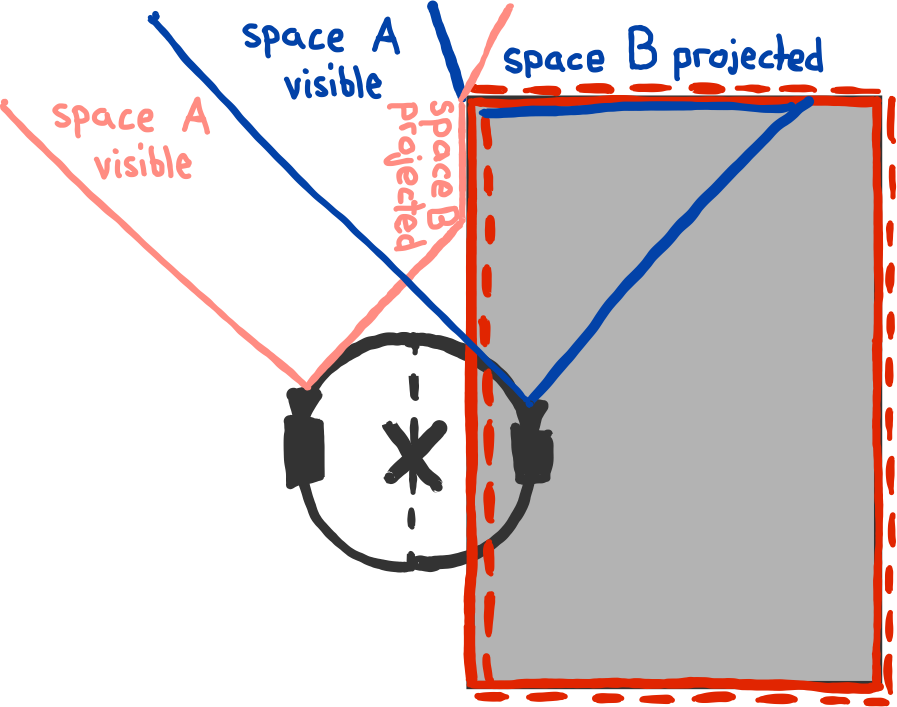}
	\caption{The sides of the portal box are only visible from the side with a solid red line. Therefore, the right eye can see space $B$ on the inner wall and simultaneously look out of the portal box back into space $A$.}
	\label{fig:unobtrusive}
\end{figure}

The final problem is that if this eye wants to look back out from the portal box into the old room, it should not see the portal plane. This is where we use back-face culling to make sure the front side of the portal box is only visible from the outside. A visualisation of what each eye sees in this scenario can be found in Figure~\ref{fig:unobtrusive}.

\section{Performance Impact}
\begin{figure}[ht]
	\includegraphics[width=\linewidth]{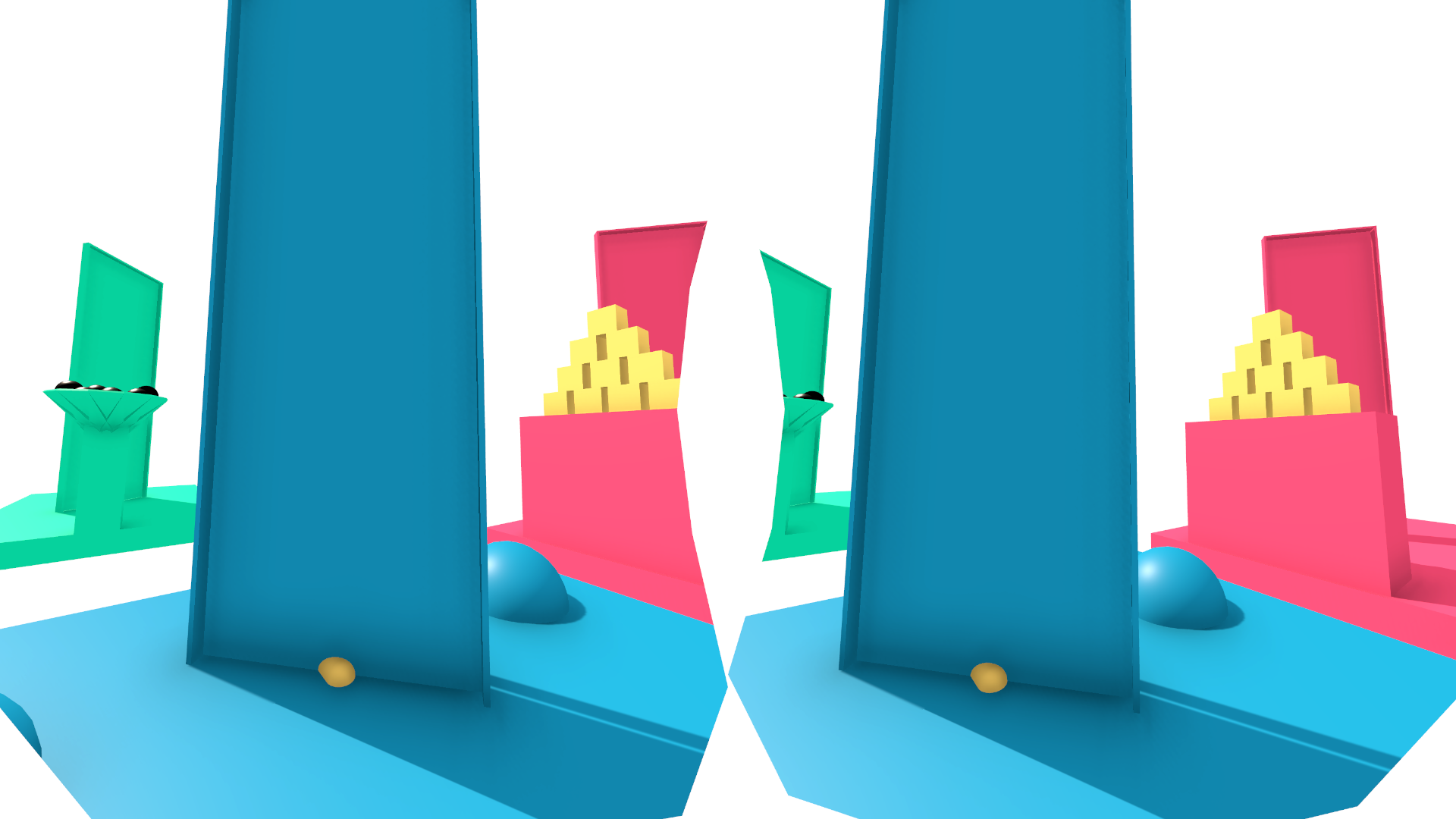}
	\caption{The test scene with no portals enabled.}
	\label{fig:test-scene-0}
\end{figure}

\begin{figure}[ht]
	\includegraphics[width=\linewidth]{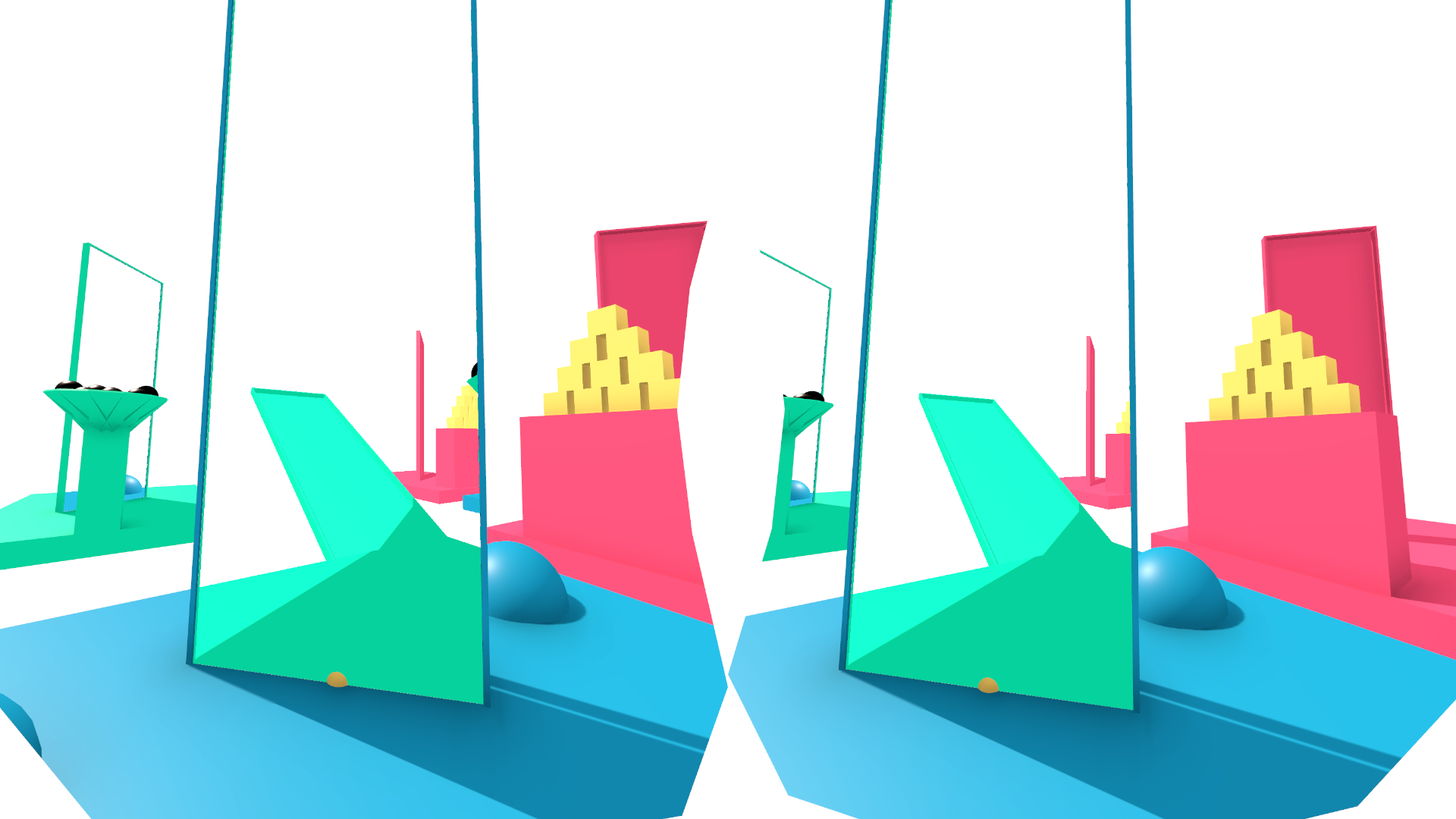}
	\caption{The connected blue and green portals are enabled.}
	\label{fig:test-scene-2}
\end{figure}

\begin{figure}[ht]
	\includegraphics[width=\linewidth]{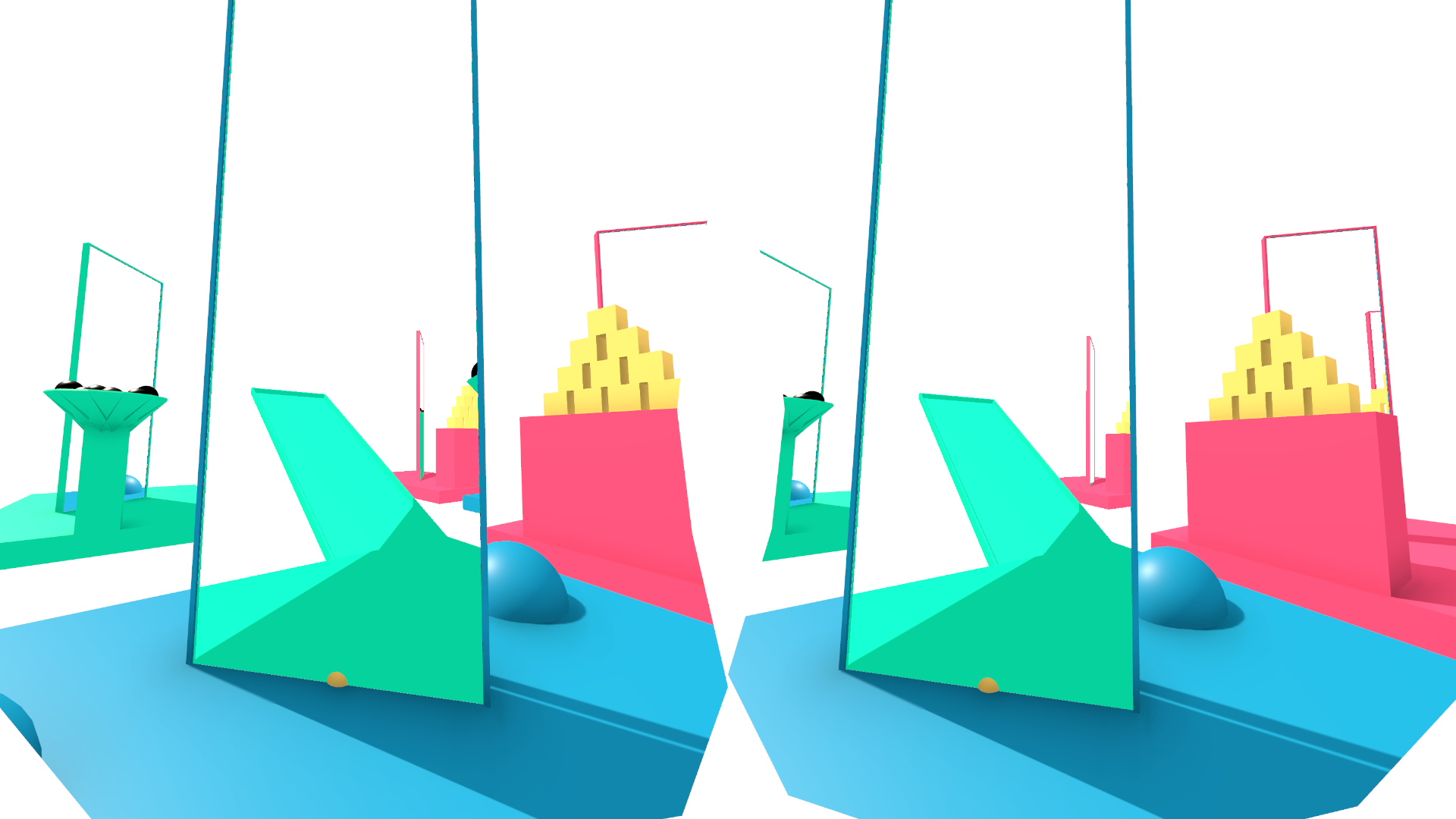}
	\caption{The test scene with four portals enabled.}
	\label{fig:test-scene-4}
\end{figure}

\begin{figure}[ht]
	\includegraphics[width=\linewidth]{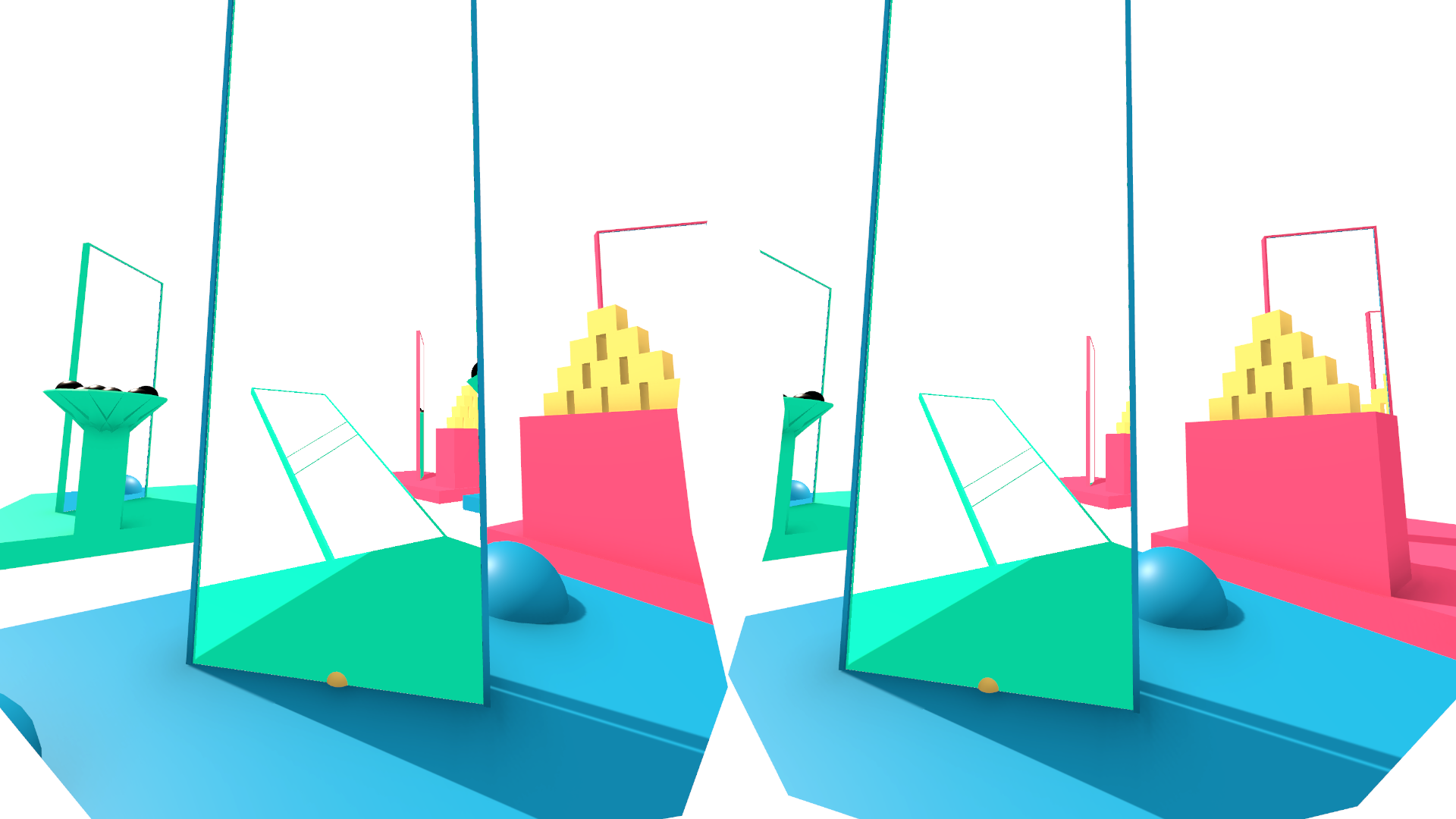}
	\caption{The test scene with all six portals enabled.}
	\label{fig:test-scene-6}
\end{figure}

To better understand the impact rendering multiple portals in VR has, we created a test application with three pairs of connected portals. With this test scene, we can measure the performance impact of rendering one, two or three pairs of portals and compare it to the baseline of no portals. An illustration of this test scene can be found in Figures \ref{fig:test-scene-0} through \ref{fig:test-scene-6}.

We measured the rendering performance of this test scene four times, each time with a different number of active portal pairs. The tests were all performed on an Nvidia GeForce GTX 1070 with the only changes between the tests being the number of active portals. The results of our tests can be found in the following table:

\vspace{5mm}
\begin{tabular}{l|l|l|l|l}
	\textbf{Portals} & \textbf{0} & \textbf{2} & \textbf{4} & \textbf{6}
	\begin{filecontents*}{assets/measurements.csv}
Portal Count,0 portals,2 portals,4 portals,6 portals
Frame Rate [fps],44.7,27.9,17.4,10.3
Frame Time [ms],22,37,58,98
Main Thread [ms],9,23,43,77
Render Thread [ms],18,24,28,40
GPU Frame [ms],0.3,14,44,78
\end{filecontents*}
\csvreader{assets/measurements.csv}{}
	{\\\hline\csvcoli&\csvcolii&\csvcoliii&\csvcoliv&\csvcolv}
	\label{tab:measurements}
\end{tabular}
\vspace{5mm}

The measured values were the average frame rate of the application in frames per second, the average time taken for the whole frame and the individual times in the main and render threads on the CPU and the GPU render thread.

\begin{center}
	\begin{figure}[ht]
		\includegraphics[width=.75\linewidth]{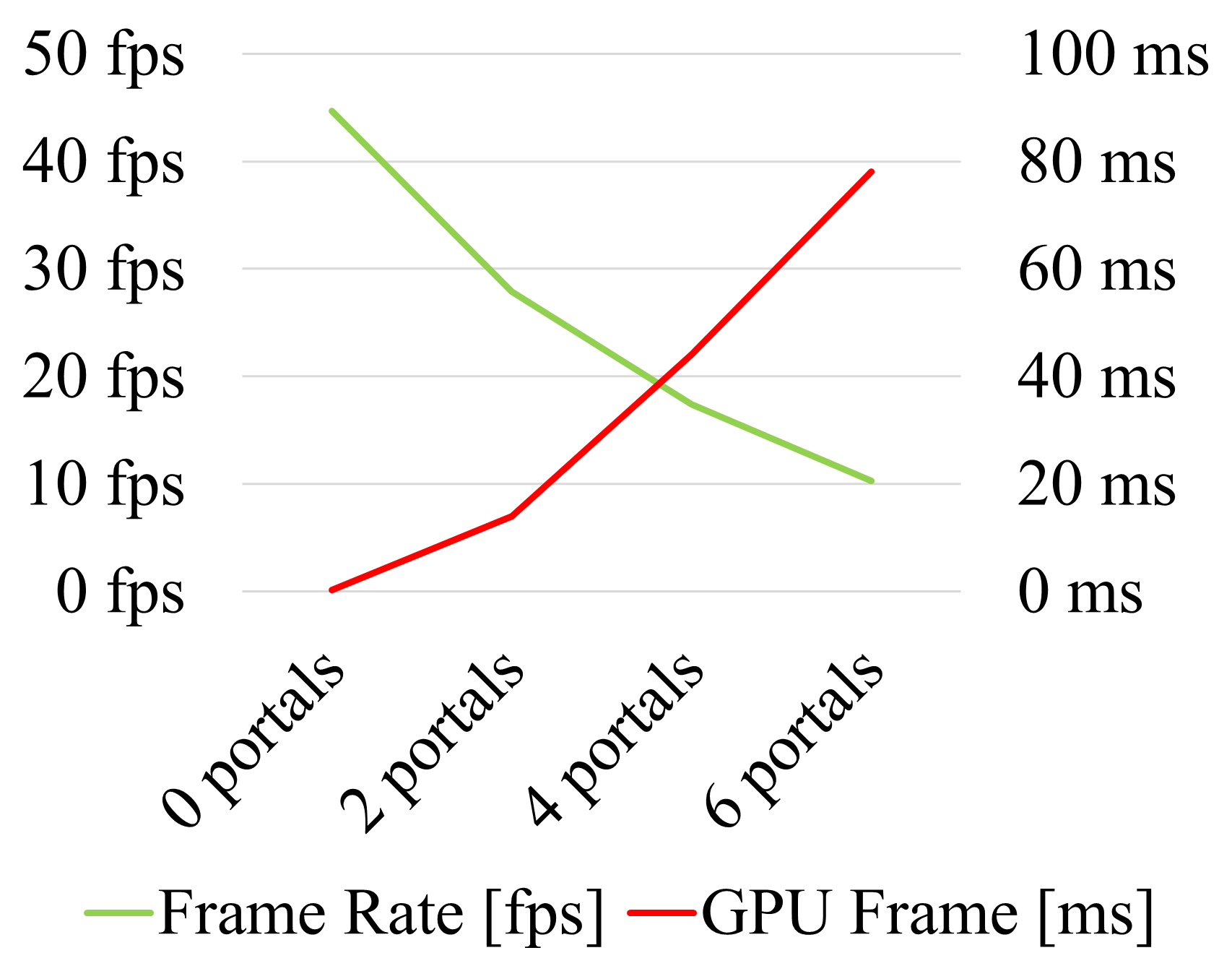}
		\caption{}
		\label{fig:measurements}
	\end{figure}
\end{center}

Since our focus is on the performance impact of rendering the portals, we focus on the frame rate and the GPU frame time. Those two measurements are visualised in Figure \ref{fig:measurements}.

We note that rendering only a single pair of portals already incurs a large cost, as seen in the decrease of the frame rate by approximately $37\%$ from $44.7$ fps down to $27.9$ fps. The almost $50$ fold increase in GPU frame time suggests that this performance loss happens mostly on the GPU. The baseline test without any portals does not use any textures. Ji et al.\cite{lod:2005} mention the performance cost of pushing texture data to the GPU, which explains the increased frame time when using textures to render the portals.

Drawing the scene without portals can be done in two render passes, one for each eye. Every added portal requires drawing the scene again from each eye, thus a single portal pair increases the number of passes from two to six. With three pairs of portals, the scene requires a total of 14 render passes. The graph in Figure \ref{fig:measurements} shows this growth of the frame time.

\section{Optimisation Considerations}

\begin{figure}[ht]
	\includegraphics[width=\linewidth]{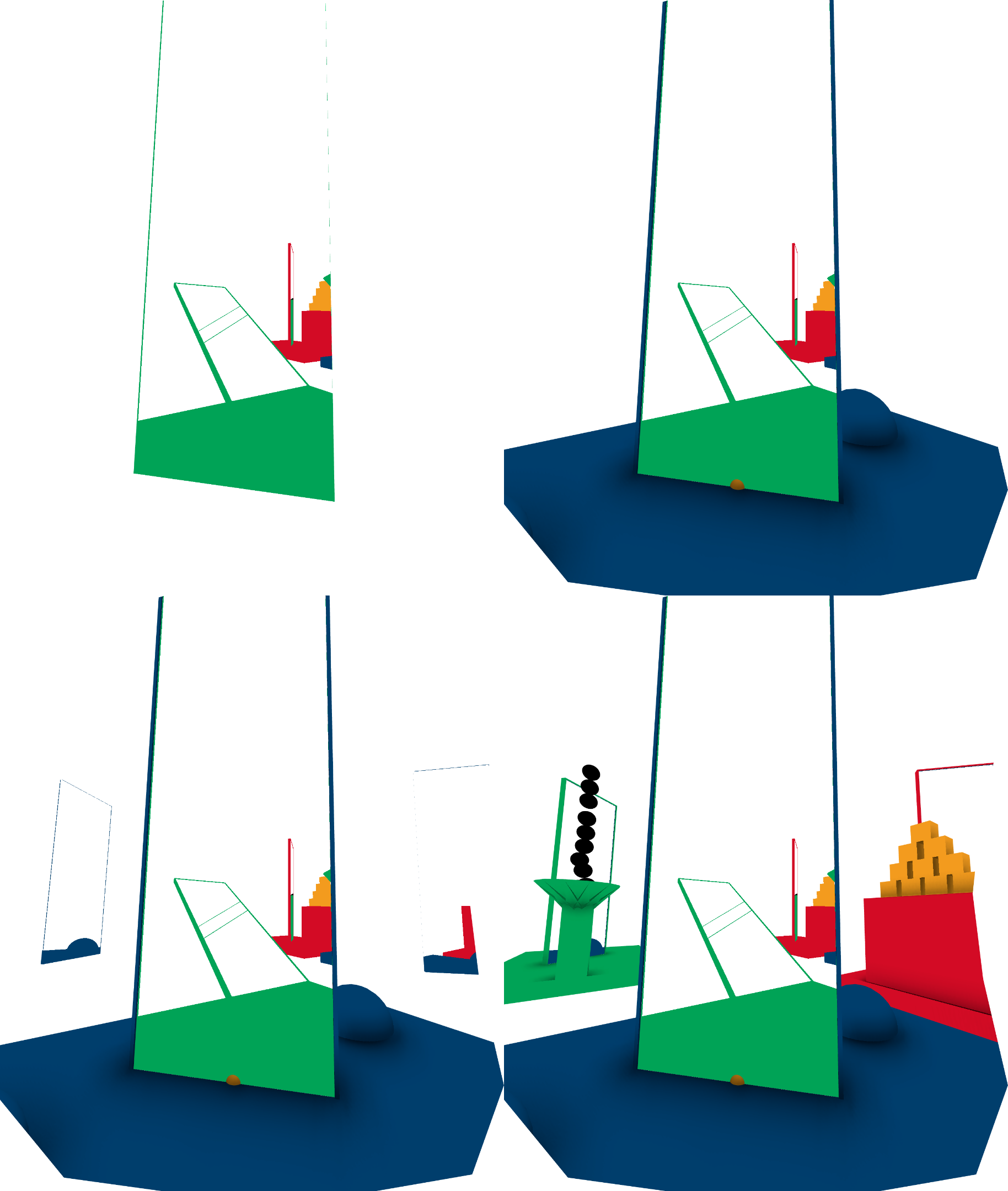}
	\caption{This figure shows how the left eye of our test scene is being rendered. It begins by rendering the view from the blue portal (top left), followed by the geometry of the blue platform (top right). Similarly, the rest of the portals are rendered first (bottom left) and the platform they are on second (bottom right).}
	\label{fig:render-steps}
\end{figure}

Rendering each portal portal requires an additional viewpoint. In VR, where each eye is rendered by its own camera, there are now two additional viewpoints to render from per portal. In the naïve case, where each viewpoint is rendered the same, we produce quadruple the amount of work compared to a monoscopic scene without portals.

Rendering the whole view from a portal can lead to a lot of wasted performance. For example, the further away a portal is, the fewer pixels on the screen it occupies. Rendering the view from the portal in full resolution can be wasteful if the portal is far from the camera or completely redundant if the user is not looking at the portal.

Furthermore, the portal itself might be blocked by other geometry in front of it. Figure \ref{fig:render-steps} shows how the two portals rendered in the bottom left image are subsequently being blocked by the green bowl and the yellow cubes in the final image.

We will present two optimisations that can reduce the rendering time and analyse their impact.

\subsection{Stencil Buffer}
The first optimisation we consider uses the so-called stencil buffer to only render what is seen through the portal. This can improve the rendering time of portals in general.

\begin{figure}[ht]
	\includegraphics[width=\linewidth]{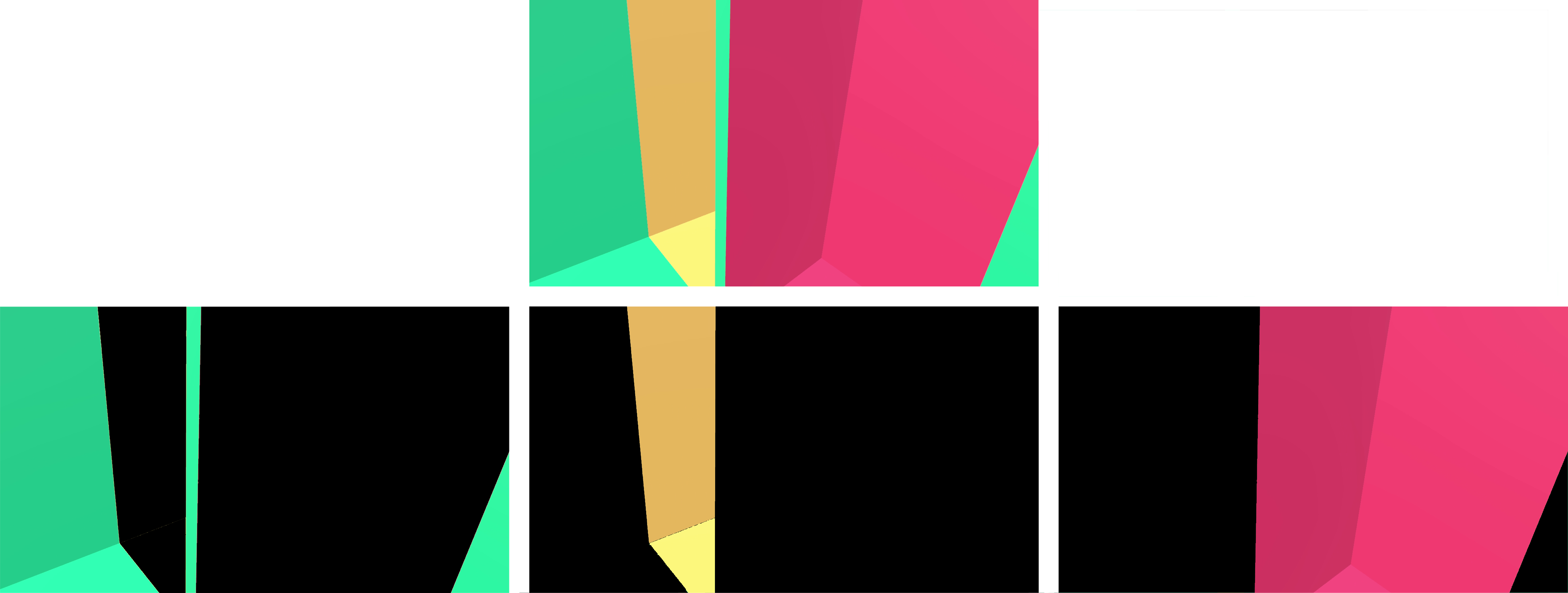}
	\caption{The user is standing in a green room looking at two portals to a yellow and a magenta room. The amount of rendering can in this case be divided by three if the pairs of cameras only render a masked out area of their room instead of the whole picture.}
	\label{fig:stencil-portals}
\end{figure}

The stencil buffer on graphics cards is used to mask out an area on the screen that will not be drawn to. If the portals are first rendered with a shader that simply marks the stencil buffer and nothing else, in a second pass, the virtual cameras behind those portals can then fill the remaining area with the view from the other side of the portals. This way, no redundant pixels are drawn and the amount of pixels that need to be drawn stays practically constant no matter how many portals are visible on the screen. An example demonstrating how a scene with two portals is split up by a stencil buffer can be found in Figure~\ref{fig:stencil-portals}.

As mentioned by Vlachos\cite{vlachos:2015}\cite{vlachos:2016}, the stencil buffer is also used to cull the pixels not visible in the VR headset. This is apparent in Figure \ref{fig:render-steps}, where a rounded off part at the bottom of the blue platform would never be visible to the user and is left out.

\subsection{Single-Pass Instanced Rendering}
In contrast to the stencil buffer, the second optimisation improves the overall performance of rendering VR by not pushing the whole scene to the GPU twice and rendering a texture for each eye, but rather rendering both eyes onto a single texture while only keeping a single copy of each rendered object in memory.

\begin{figure}[ht]
	\includegraphics[width=\linewidth]{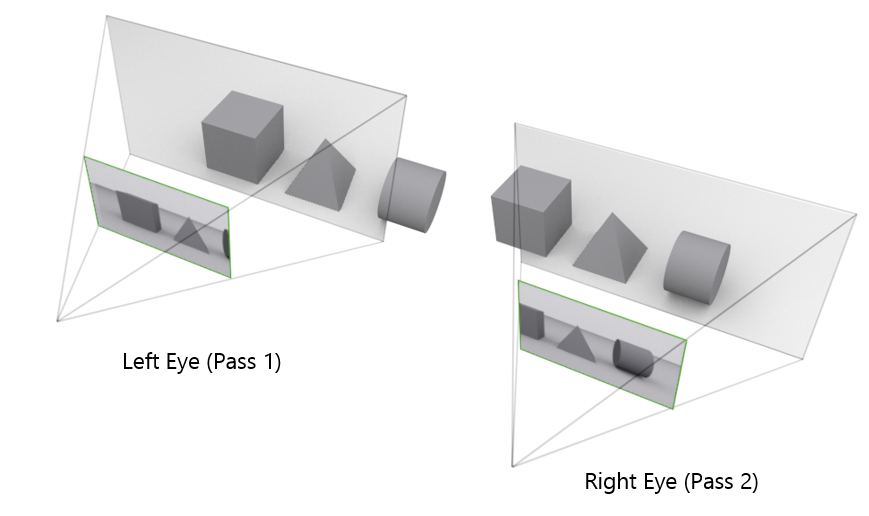}
	\caption{When rendering in multiple passes, all work is doubled. This illustration shows how basically the three objects are treated as though they were separate objects\cite{nvidia-sps}.}
	\label{fig:multi-pass}
\end{figure}

The naïve way of rendering a stereoscopic image is to simply render it two times, once from the perspective of each eye. This is called multi-pass rendering and is illustrated in Figure~\ref{fig:multi-pass}. The advantage of this technique is its simplicity in implementing it. Shader code from monoscopic applications can be used in multi-pass rendering without any changes since we are essentially just rendering two images per frame.

This method comes with a large performance downside. In the worst case, the amount of rendering work is multiplied by the number of portals. Thus, the rendering overhead caused by multi-pass rendering is in this case not acceptable.

\begin{figure}[ht]
	\includegraphics[width=\linewidth]{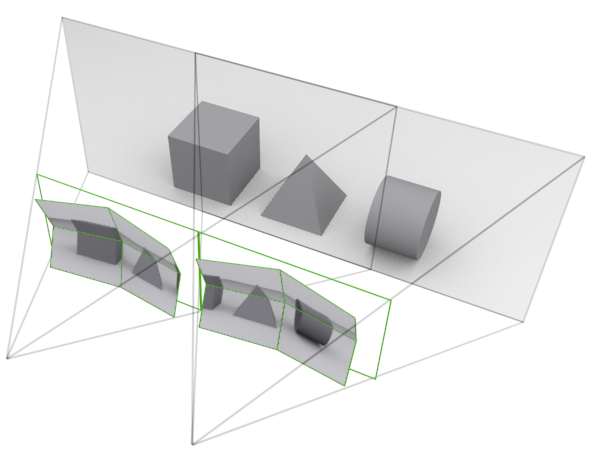}
	\caption{When rendering in an instanced single pass, the whole geometry of the scene is processed only once, which significantly reduces the amount of work\cite{nvidia-sps}.}
	\label{fig:single-pass}
\end{figure}

The second method described in Figure~\ref{fig:single-pass} --- single-pass instanced rendering --- improves upon this by rendering both the left and right eyes onto the same texture. Additionally, the geometry only exists on the GPU once, as well as any shadow maps and other view-independent data. The objects are rendered twice as different instances from the two perspectives of the eyes.

Single-pass instanced rendering should greatly improve the performance, but as mentioned in Section~\ref{unobtrusive-transitions} rendering both eyes in one pass eliminates the possibility of transporting them through the portal separately from each other.

\section{Conclusion}

We have shown how to implement portals in VR that allow for an unobtrusive transition through them. This can help immersion in an impossible space of artificially increased size.

The tests with our implementation of portals shows that it is very expensive to render each pair of portals. Currently, a scene should be limited to having as few portals visible as possible. In an experience where the user progresses through a series of rooms in a linear fashion, the amount of portals could be limited by disabling portals of previously visited rooms as soon as the next one is entered.

The mentioned performance optimisations might help making a VR experience that has many portals visible at the same time possible. Using the stencil buffer to render only the necessary parts of each portal could in theory allow for an unlimited amount of portals shown at the same time. We showed that the amount of pixels rendered in this case stays practically constant.

\bibliographystyle{plain}
\bibliography{refs}

\begin{thebibliography}{10}
\providecommand{\url}[1]{#1}
\csname url@samestyle\endcsname
\providecommand{\newblock}{\relax}
\providecommand{\bibinfo}[2]{#2}
\providecommand{\BIBentrySTDinterwordspacing}{\spaceskip=0pt\relax}
\providecommand{\BIBentryALTinterwordstretchfactor}{4}
\providecommand{\BIBentryALTinterwordspacing}{\spaceskip=\fontdimen2\font plus
\BIBentryALTinterwordstretchfactor\fontdimen3\font minus
  \fontdimen4\font\relax}
\providecommand{\BIBforeignlanguage}[2]{{%
\expandafter\ifx\csname l@#1\endcsname\relax
\typeout{** WARNING: IEEEtran.bst: No hyphenation pattern has been}%
\typeout{** loaded for the language `#1'. Using the pattern for}%
\typeout{** the default language instead.}%
\else
\language=\csname l@#1\endcsname
\fi
#2}}
\providecommand{\BIBdecl}{\relax}
\BIBdecl

\bibitem{lowe:2005}
N.~Lowe and A.~Datta, ``A new technique for rendering complex portals,''
  \emph{IEEE Transactions on Visualization and Computer Graphics}, vol.~11,
  no.~1, pp. 81--90, 2005.

\bibitem{aliaga:1997}
D.~G. Aliaga and A.~A. Lastra, ``Architectural walkthroughs using portal
  textures,'' in \emph{Proceedings. Visualization '97 (Cat. No.
  97CB36155)}.\hskip 1em plus 0.5em minus 0.4em\relax IEEE, 1997, pp. 355--362.

\bibitem{abrash:2013}
\BIBentryALTinterwordspacing
M.~Abrash, ``Why virtual reality is hard (and where it might be going),''
  \emph{Game Developers Conference 2013}, 2013. [Online]. Available:
  \url{https://www.valvesoftware.com/en/publications}
\BIBentrySTDinterwordspacing

\bibitem{boletsis:2017}
\BIBentryALTinterwordspacing
C.~Boletsis, ``The new era of virtual reality locomotion: A systematic
  literature review of techniques and a proposed typology,'' \emph{Multimodal
  Technologies and Interaction}, vol.~1, no.~4, 2017. [Online]. Available:
  \url{https://www.mdpi.com/2414-4088/1/4/24}
\BIBentrySTDinterwordspacing

\bibitem{lochner:2021}
D.~Lochner, ``Vr natural walking in impossible spaces,'' \emph{Motion,
  Interaction and Games (MIG '21)}, 2021.

\bibitem{monoscopic-stereoscopic:2020}
M.~B. Taştı and Ümmühan Avcı, ``Examination of using monoscopic
  three-dimensional (m3d) and stereoscopic three-dimensional (s3d) animation on
  students,'' \emph{Education and Information Technologies}, vol.~25, pp.
  2765--2790, 2020.

\bibitem{back-face-culling:2007}
D.-B. Jeon, S.-Y. Kim, K.-Y. Lee, and J.-C. Kwak, ``A design of a 3d graphics
  rasterizer with a culling and clipping,'' in \emph{TENCON 2007 - 2007 IEEE
  Region 10 Conference}, 2007, pp. 1--4.

\bibitem{lod:2005}
J.~Ji, E.~Wu, S.~Li, and X.~Liu, ``Dynamic lod on gpu,'' in \emph{Proceedings
  of the Computer Graphics International 2005}, ser. CGI '05.\hskip 1em plus
  0.5em minus 0.4em\relax USA: IEEE Computer Society, 2005, p. 108–114.

\bibitem{vlachos:2015}
\BIBentryALTinterwordspacing
A.~Vlachos, ``Advanced vr rendering,'' \emph{Game Developers Conference 2015},
  2015. [Online]. Available:
  \url{https://www.valvesoftware.com/en/publications}
\BIBentrySTDinterwordspacing

\bibitem{vlachos:2016}
\BIBentryALTinterwordspacing
------, ``Advanced vr rendering performance,'' \emph{Game Developers Conference
  2016}, 2016. [Online]. Available:
  \url{https://www.valvesoftware.com/en/publications}
\BIBentrySTDinterwordspacing

\bibitem{nvidia-sps}
NVIDIA, ``Vrworks -- single pass stereo,''
  \url{https://developer.nvidia.com/vrworks/graphics/singlepassstereo},
  [accessed 1.11.2022].

\end{thebibliography}

\end{document}